\documentstyle[12pt,epsfig]{article}
\begin{document}
\begin{titlepage}

\begin{center}
  {\Large \bf{Simultaneously non-linear energy calibration of CMS calorimeters for single pions and electrons}}
\end{center}

\vspace{10mm}

\begin{center}
    J. Damgov$^{*}$, V. Genchev$^{*}$, S. Cht. Mavrodiev

       INRNE, Sofia, Bulgaria
\end{center}

\begin{center}
\today
\end{center}

\vspace{10mm}

\begin{abstract}
 CMS calorimeter energy calibration was done in the full CMS simulated 
 geometry for the pseudorapidity region $\eta$ = 0. The samples of
 single pion events were generated with a set of incident energies from 5 
 GeV to 3 TeV and for single electrons from 5 to 500 GeV. The analysis 
 of the simulated data shows that standard calibration using just sampling 
 coefficients for calorimeter parts with different sampling ratio gives 
 nonlinear calorimeter response. Non-linear calibration technique was 
 applied simultaneously for pion and electron beams which is preparation for
 jets energy reconstruction. It improve calorimeter energy resolution for 
 pions and restore the calorimeter linearity.
\end{abstract} 

\vspace{90mm}

$^{*}$This study is supported by the Bulgarian Ministry of Education and Sciences
  
\end{titlepage}

\section{Calorimeters geometry considered}

 The pion and electron responses are obtained with GEANT simulations with a 
detailed description of CMS calorimeter geometry (CMSIM 115), 
version TDR-2$^{/1/}$.

The ECAL is the PbWO$_{4}$ one (readout number 1).

An additional 1 cm scintillator layer is placed in front of HB
to compensate for the energy loss in the cables, electronics and cooling 
system on the back side of the ECAL, which is simulated by a uniform slab 
of 0.2 $\lambda$ of dead material (readout number 2).
 
The sampling thicknesses of copper alloy (90\%Cu + 10\%Zn) are 5 cm in the 
barrel segments, except the inner and outer plates 
which are 7 cm stainless steel. The 9 mm gaps between the absorber plates 
are filed with 4 mm scintillator planes (readout number 3). 

To improve the energy measurement for late developing hadron showers, 
tail catcher layers are inserted in front of and behind the first return 
yoke iron layer (RY1) (readout number 4).  
 
The regions behind the ECAL are empty except for 
the scintillator layer and equivalent dead material uniform absorber layer. 
The effects of the CMS 4T field are fully included. 

\section{Data sample}

We have performed a GEANT simulation of the response to pions for a set 
of incident energies from 5 to 3000 GeV  and to electrons from 5 to 
500 GeV at pseudorapidity set to 0. GHEISHA was used as a hadron interaction 
simulator. The sum of the energies deposited by the showers in the active 
elements of each readout has been stored 
in a disk file so that the showers could be analysed later. 

\section{Standard calibration}

The standard calibration uses sampling coefficients for calorimeter 
parts with different sampling ratio. 
The reconstructed energy $ E^{rec}$ of simple shower is given by the 
weighted sum of the energies deposited in the readouts:

 $$ E^{rec} = \sum_{i=1, 4} c_{i} E_{i} , $$

where:

$E_{i}$ - amplitude of the signal from the calorimeter longitudinal 
segmentation (readouts);

$c_{i}$ - calibration coefficients, determined by the minimisation of 
the width of the energy distributions.

The Gaussian part of the reconstructed energy distributions at various incident 
energies are then fitted to obtain the calorimeter energy 
resolution. The energy resolution is parametrised by the expression:

$$ \sigma /E = a/ \sqrt{E} \oplus b. $$ 

The energy resolution obtained by the standard calibration with 
simultaneously fit of pion and electron events is shown in fig. 1a) and 1b) 
and the residuals of the reconstructed energy in fig. 2a) and 2b) 
(open points).    

\section{Non-linear technique}

The linear behaviour of the calorimeter response could be restored by 
application of the non-linear method which improve 
the linearity of the calorimeter response and the energy resolution in the 
broad energy range.
Non-linear technique is the selection of some additional energy depended 
parameters which provide correct energy reconstruction of the showers.

Thus the reconstructed energy $E^{rec}_{nl}$ is parametrised as:

 $$ E^{rec}_{nl} = \sum_{i=1, 4} f_{i}(\vec{A}, E_{i}) E_{i} , $$

where $f_{i}$ are non-linear functions of N unknown parameters 
$A_{j}$, j = 1, N and measured readout signals $E_{i}$. 

The system is solved with autoregularized Newton type 
method$^{/2/}$ by minimisation of the $\chi^{2}$ functional (REGN code$^{/3/}$)

 $$ \chi^{2} = \sum_{j=1, M} W_{j}(E_{j}^{in} - E_{nl, j}^{rec})^{2},$$

where M is number of generated set of energies multiplied by the number 
of generated events at each energy and $W_{j}$ are statistical weights.

In the REGN computer code the $\chi^{2}$ is one 
of the different criteria available for solving of the system and for 
testing the mathematical model. The other criteria permit one to chose 
uniquely between several model functions the best one$^{/4, 5, 6/}$.

This technique was applied for single pions calibration$^{/7/}$. 
The obtained energy resolution after applying the non-linear calibration 
method for simultaneously fit of pion and electron beams is shown on fig. 1a)
and 1b) (full points). Resolution for pions becomes better in comparison 
to standard linear calibration. A small degradation of electron energy 
resolution is observed. The comparison of standard calibration method to the 
non-linear technique gives a significant improvement in the linearity of 
the reconstructed energy for both pions and electrons (fig. 2a) and 2b))
(full points). 

The unification of our technique with the most clear physical 
approaches$^{8, 9, 10}$ can give better energy resolution because we can 
test strait different energy dependences of unknown parameters which 
depends of interaction cross-sections and calorimeters construction details.

\section{Summary} 
 
The CMS calorimeters simultaneously pion and electron calibration was done 
using linear and non-linear approaches. The standard calibration method 
gives huge deviation from linearity of the reconstructed energy. The
calibration using non-linear technique improves behaviour of the energy 
resolution for pions and restores the calorimeter linearity. Such 
non-linear technique can be tested successfully for pions, electrons and 
jets cases simultaneously.

%\newpage

\vspace{30mm}

\begin{figure}[hbtp]
  \begin{center}
    \resizebox{14cm}{!}{\includegraphics{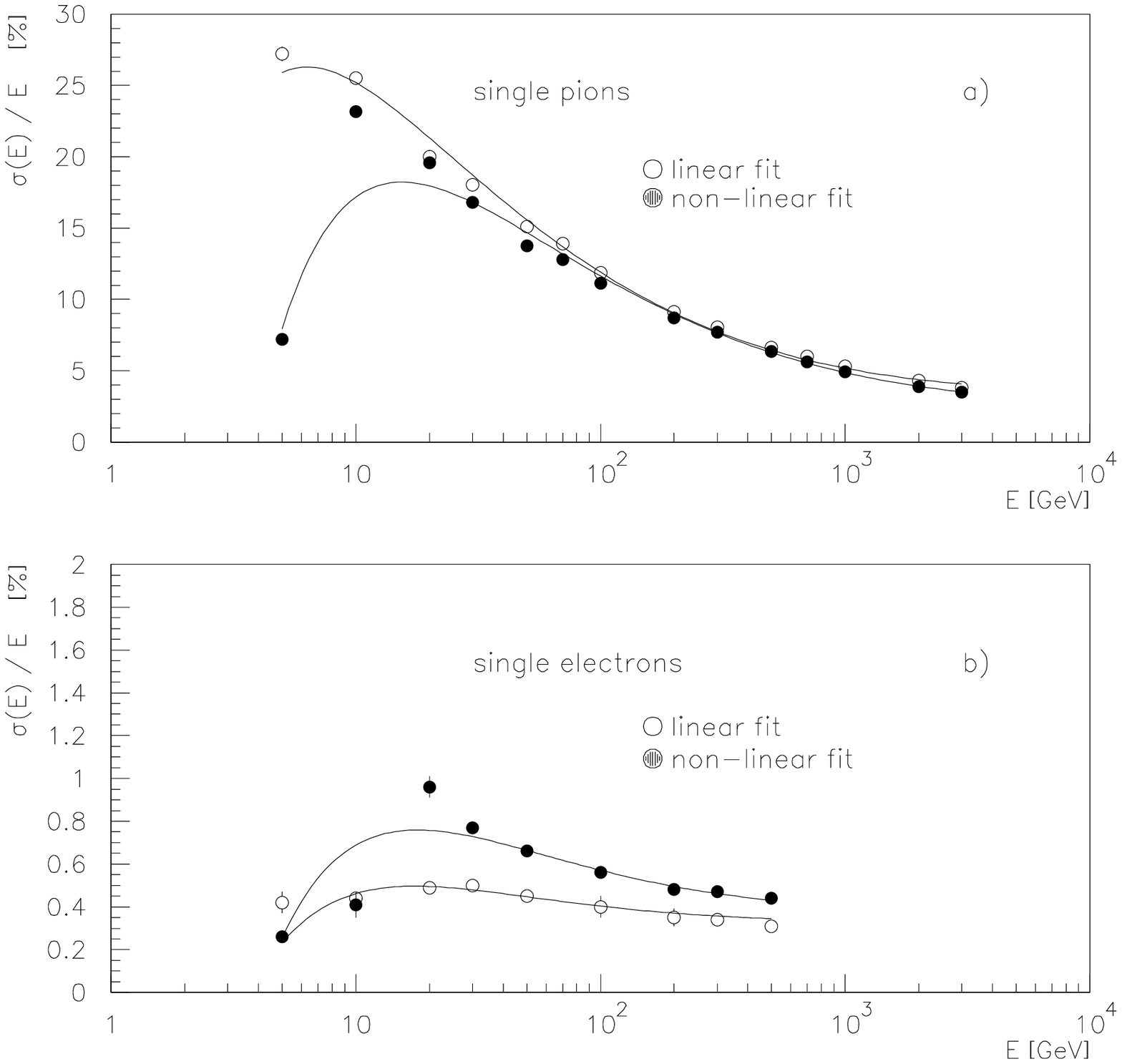}}
\caption{Energy resolution for the CMS calorimeter system (standard
calibration - open points, non-linear calibration - full points).} 
\label{fig:1}
  \end{center}
\end{figure}

\begin{figure}[hbtp]
  \begin{center}
    \resizebox{14cm}{!}{\includegraphics{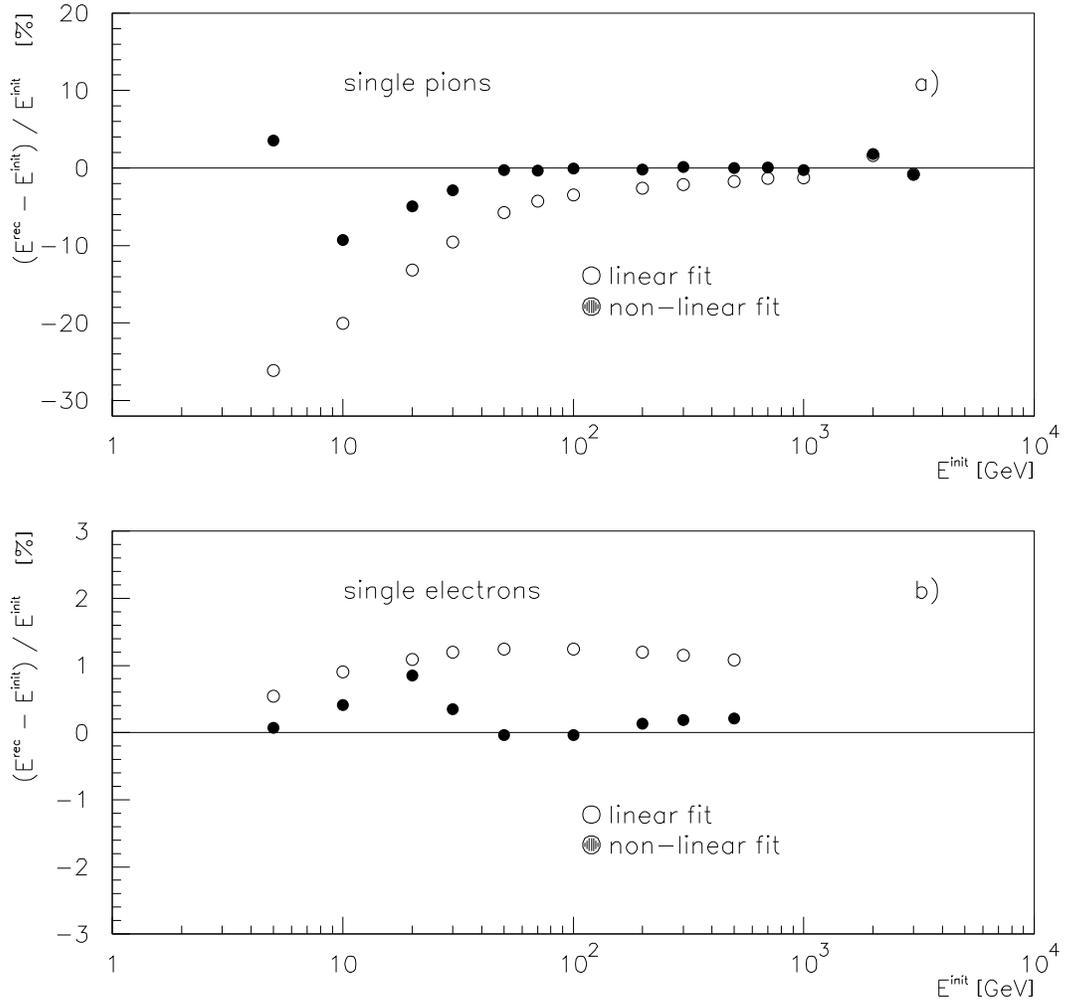}}
\caption{Residuals of the reconstructed energy (standard
calibration - open points, non-linear calibration - full points).} 
\label{fig:2}
  \end{center}
\end{figure}

\end{document}